\documentclass[twocolumn, amssymb, nobibnotes, aps, prl, superscriptaddress, floatfix, epsfig, graphics]{revtex4}

\usepackage{graphicx}
\usepackage{amsmath}
\usepackage{dcolumn}
\usepackage{bm}
\usepackage{accents}

\begin{document}

\title{Ising Model for the Freezing Transition}
\author{Jacobo Troncoso}
\email{jacobotc@uvigo.es}
\author{Claudio A. Cerdeiri\~na}
\email{calvarez@uvigo.es}
\affiliation{Departamento de F\'isica Aplicada and Instituto de F\'isica y Ciencias Aeroespaciales, Universidad de
  Vigo---Campus del Agua, Ourense 32004, Spain} 

\date{\today}

\begin{abstract}

A spin-1 Ising model incorporating positional order to a standard lattice gas with no attractive interactions is introduced and found to be consistent with all known attributes of the freezing transition of the hard-sphere system. Implementation of attractive interactions in a fairly natural way then allows every aspect of the phase diagram of a simple substance to be reproduced. The \emph{whole} phase behavior of such sort of substances is thus found to sharply manifest the van der Waals picture highlighting the relevance of harsh repulsive forces.

\noindent
\end{abstract}

\maketitle

\noindent

An outstanding question to the theory of condensed matter is the statistical-mechanical characterization of the freezing transition of a system of hard spheres in three dimensions \cite{alder}. Up to date, there is neither a rigorous proof of its existence nor even ``good heuristics'' to approach the problem theoretically \cite{lebowitz}. The topic is relevant to diverse problems in Physics and Mathematics \cite{parisizamponi} and has received a great deal of attention as a most basic example of entropy-driven phase transition \cite{frenkel,sciortino}, while it is thought since long ago \cite{longuet-higginswidom} to underlie the solid-liquid transition of real substances.

Here we show that a spin-1, three-state Ising model pertaining to the ``Blume-Capel'' or ``Blume-Emery-Griffiths'' class \cite{blume,capel,beg} exhibits a transition akin to the freezing of hard spheres revealed by molecular simulation. We find it crucial to our approach the notion of \emph{compressible cells} and \emph{locally fluctuating free volumes} introduced by Fisher and coworkers a few years ago \cite{yangyang}. After describing the model for the primitive hard-sphere system, we show that incorporation of attractive interactions leads to all known basic features of the phase diagram of a simple monatomic substance like, e.g., argon. We work at the mean-field level throughout. The manuscript ends with a few concluding remarks.

\emph{Hard-sphere system.}---Consider the three-dimensional space divided into elementary rhombic dodecahedra, as Wigner-Seitz primitive cells of an underlying fcc regular lattice \cite{ashcroftmermin}. As Fig. 1a illustrates, in two of the  three states for an individual cell it can be empty or singly occupied in a ``disordered'' configuration by a hard-sphere particle with diameter $\sigma$ whose center explores a free volume $\dot v_0$ in it. We may safely express the cell volume as $\lambda \sigma^3$ with $\lambda$ sufficiently large, while the $\dot v_0\leq\lambda \sigma^3$ constraint adds to the condition of forbidden multiple occupancy to account for harsh repulsive forces. A model with these two states for cells is in essence a standard lattice gas \cite{eyring,yanglee} with no attractive interactions.

\begin{figure}[!h] 
\begin{center}
          \includegraphics[width=9 cm]{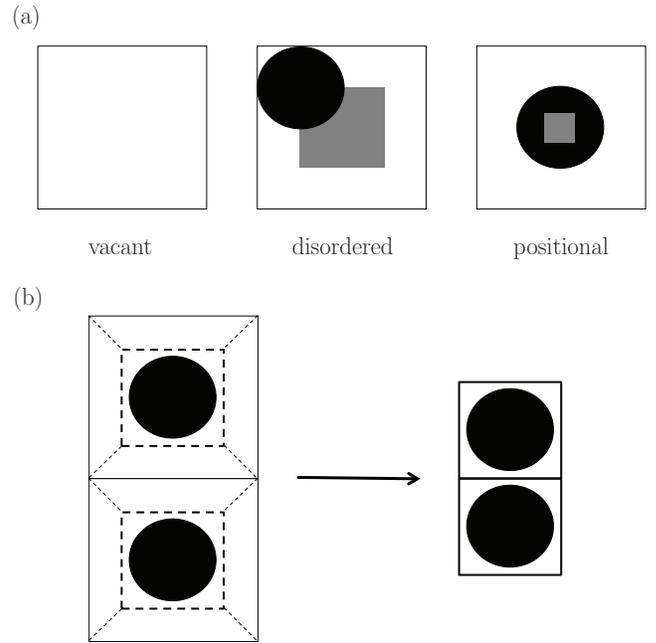}
           \caption{Two-dimensional illustration of our hard-sphere model for a square lattice. (a) Individual cell states, with the shaded grey areas representing the free volume explored by particles in ``disordered'' and ``positional'' states ($\dot v_0$ and $\dot v_1$, respectively). (b) Assembly of two nearest-neighbor cells with particles with positional order in the hypothetical case that all their six nearest neighbors likewise contain particles with positional order. Accordingly, both cells receive inputs delimitated by the isosceles trapezoids in the left graph to become the squares of reduced area in the right one.}
			\label{fig1}
\end{center}
\end{figure}

Now, in addition to the disordered and vacant states, we suppose that there is a third, \emph{special} state in which a particle just explores a preferential, restricted free volume $\dot v_1<\dot v_0$ around the center of its cell (see Fig. 1a). The mechanism of molecular packing is then implemented on simply postulating that the total volume of an assembly of two nearest-neighbor cells is decreased when particles in them have the prescribed \emph{positional order}. This is illustrated by Fig. 1b, where we further assume that the volume decrease is equally shared between neighboring cells. Under these circumstances, we are led to imagine the picture that a cell with positional order and with all of its nearest neighbors likewise in that same state is a smaller rhombic dodecahedron whose decreased volume equals the fcc volume per particle at close packing, namely, ${\scriptstyle\frac{\sqrt 2}{2}} \sigma^3$ \cite{closepacking}. Fulfillment of this limiting condition is imposed by fixing the two-cell volume decrease to ${\scriptstyle\frac{1}{6}}(\lambda-{\scriptstyle\frac{\sqrt 2}{2}})\sigma^3$ \cite{comment}. There is certain degree of artificiality inasmuch as most accessible microstates $l$ may contain cells with distict size and shape that are inconsistent with a continuum-space model, but this shall be of no major consequence.

A statistical-mechanical treatment entails summing Boltzmann factors $e^{-\beta(E_l+pV_l-\mu N_l)}$ over $\{l\}$, where $E$ denotes the energy, $V$ the volume, $N$ the number of particles, $p$ the pressure, and $\mu$ the chemical potential, while $\beta\equiv 1/k_B T$ with $k_B$ the Boltzmann constant and $T$ the temperature. This embodies an ensemble in which $E$, $V$, and $N$ are allowed to fluctuate simultaneously, as introduced long ago \cite{guggenheim,sack} and being increasingly used over the years \cite{singularityfree,noonorkoulas,yangyang,isingparadigm,latella}. While one may further proceed by introducing spin-1 variables and ask for an exact solution, the task becomes rather straightforward in mean-field approximation.

Explicitly, on considering the number of cells with particles with positional order $N_+$, one finds that for a total number of cells ${\cal N}$ the following order parameters
$$
n\equiv {N\over {\cal N}},\,\,\,\,\,\,\,\,\,\, n_+\equiv {N_+\over {\cal N}}
\eqno (1)
$$
allow to write
$$
E={\scriptstyle\frac{3}{2}}n{\cal N}k_BT
\eqno (2)
$$
and
$$
V={\cal N}[\lambda- (\lambda-{\scriptstyle\frac{\sqrt 2}{2}})n_+^2]\sigma^3.
\eqno (3)
$$
On the other hand, the entropy $S$ splits into a configurational contribution $S_{\rm conf}$ and a free volume one $S_{\rm fv}$ given by
$$
{S_{\rm conf}\over {\cal N}k_B}=-[n_+\ln n_+ +(n-n_+)\ln (n-n_+)+(1-n)\ln (1-n)]
\eqno (4)
$$
and
$$
{S_{\rm fv}\over {\cal N}k_B}=[n\ln (\dot v_0\Lambda_T^{-3})+n_+\ln \omega],
\eqno (5)
$$
where $\omega = \dot v_1/\dot v_0<1$ while $\Lambda_T\equiv h/\sqrt {2\pi mk_BT}$ stands for the de Broglie thermal wavelength for a particle of mass $m$, with $h$ the Planck constant. 

The number density $\rho\equiv N/V$ is readily obtained from (1) and (3), so that we may write 
$$
\bar\rho={n\over \lambda-(\lambda -{\scriptstyle\frac{\sqrt 2}{2}})n_+^2},
\eqno (6)
$$
with $\bar\rho\equiv \rho\sigma^3$. On the other hand, the standard minimization procedure leads to \cite{meanfield}
$$
{\lambda+(\lambda -{\scriptstyle\frac{\sqrt 2}{2}}) n_+^2\over 2(\lambda -{\scriptstyle\frac{\sqrt 2}{2}}) n_+}={\ln(1-n)\over\ln[\omega(n-n_+) n_+^{-1}]},
\eqno (7)
$$
$$
\bar p =-{\ln(1-n)\over \lambda+(\lambda -{\scriptstyle\frac{\sqrt 2}{2}})n_+^2},
\eqno (8)
$$
with $\bar p\equiv p\sigma^3/k_BT$. Equations (6) to (8) provide an implicit $p\rho T$ relation mediated by $n$ and $n_+$. Note that (6) and (7) jointly indicate that only one among $\rho$, $n$, and $n_+$ is independent, implying that (8) is in practice of the form $p=k_BTf(\rho)$ as Statistical Mechanics demands for a hard-sphere system \cite{monson}.

\begin{figure}[h] 
\begin{center}
          \includegraphics[width=8 cm]{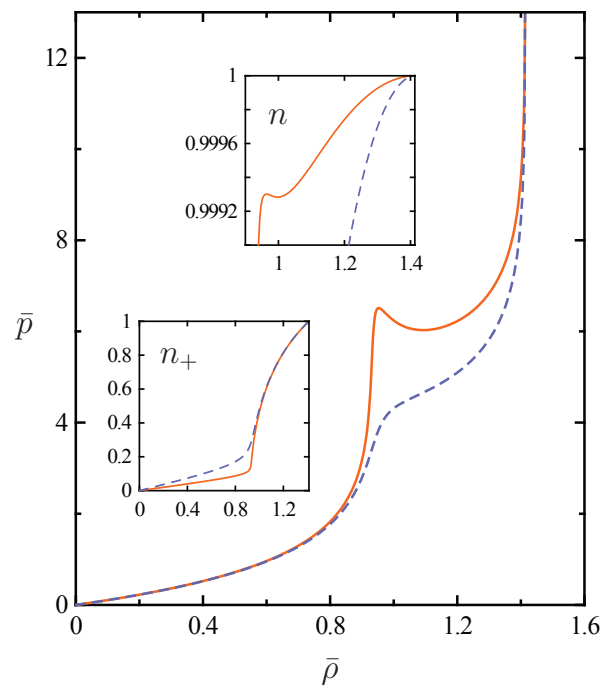}
           \caption{(Color online) Hard-sphere model values for $\bar p\equiv p\sigma^3/k_BT$ as a function of $\bar\rho\equiv \rho\sigma^3$. The insets show the order parameters $n$ and $n_+$, as a function of $\bar \rho$ too. Orange solid curves correspond to $\omega=0.1$ and purple dashed ones to $\omega=0.2$, with $\lambda =1.08$ in both cases. While not shown explicitly, $n(\bar\rho)$ is for $\omega=0.2$ a monotonically increasing curve over the whole $\bar\rho$ range.}
			\label{fig2}
\end{center}
\end{figure}

To explore the model's $\bar p(\bar\rho)$ behavior, we fix the value of $\bar\rho$ and solve (6) and (7) numerically for $n$ and $n_+$ to then use (8) to compute $\bar p$. Figure 2 shows that a van der Waals loop appears at high $\bar\rho$ when $\omega$ is sufficiently small. This is known to be the signature of a phase transition which we identify with freezing, with a customary Maxwell equal-area construction leading to a corrected curve describing coexistence. The transition also appears when other lattices such as sc and bcc are alternatively considered. Our choice, recall, a fcc lattice, is however a reasonable one since it corresponds to the hard-sphere solid that comes up with freezing.

Also evident from Fig. 2 is that $n$ is quite close to 1 throughout the loop. It is then clear from (8) that tiny variations of $n$ may be of major consequence to $\bar p$ and, certainly, inspection of Fig. 2 makes it evident that the $\bar p(\bar\rho)$ loop is mostly reflecting the behavior of $n$. Note that the concentration of vacancies, $1-n$, can be obtained from (8) to get an analytical expression, $1-n=e^{-\bar p[\lambda+(\lambda-{\scriptstyle\frac{\sqrt 2}{2}})n_+^2]}$, of the standard form \cite{ashcroftmermin}. This yields $1-n\approx 0.0003$ for the crystal at melting, which, as far as orders of magnitude are concerned, is in accord with estimations from simulations \cite{defectsalder,defectsfrenkel}.

We have verified that the spin-${\scriptstyle\frac{1}{2}}$ variant our spin-1 model reduces to for $n = 1$ leads to \emph{no} $\bar p(\bar\rho)$ \emph{loop}. The net result is that two coupled order parameters are needed to reproduce the freezing transition of the hard-sphere system: the transition arises as soon as one considers jointly monovacancies and a packing mechanism with the aid of $n$ and the positional order parameter $n_+$, respectively.

Furthermore, on calling the coexisting phases ``solid'' and ``fluid'', $n\approx 1$ allows to write $\Delta n_+\equiv n_+^{\rm solid}-n_+^{\rm fluid}\approx {\scriptstyle\frac{1}{2}}$ owing to the trivial symmetry of the effectively underlying spin-${\scriptstyle\frac{1}{2}}$ Ising model. Then, (1), (4), and (5) yield $\Delta S_{\rm conf}/Nk_B\approx 0$ and $\Delta S_{\rm fv}/Nk_B\approx {\scriptstyle\frac{1}{2}}\ln \omega$. Accordingly, $\omega=0.1$ was chosen so as to meet the hard-sphere $\Delta S/Nk_B\simeq -1.16$ value determined from molecular simulation \cite{stillinger}, while an optimal $\lambda=1.08$ value yielded the simulated $\bar\rho_{\rm fluid}\simeq 0.938$ result \cite{stillinger,haro}. These values of $\omega$ and $\lambda$ lead to $\bar\rho_{\rm solid}\simeq 1.186$, which departs from the 1.037 simulated one \cite{stillinger,haro}. One may think about further refining the model so as to get a closer numerical estimate of $\bar\rho_{\rm solid}$, but such an expediency is beyond the scope of the present discussion \cite{comment2}.

\begin{figure}[h] 
\begin{center}
          \includegraphics[width=8 cm]{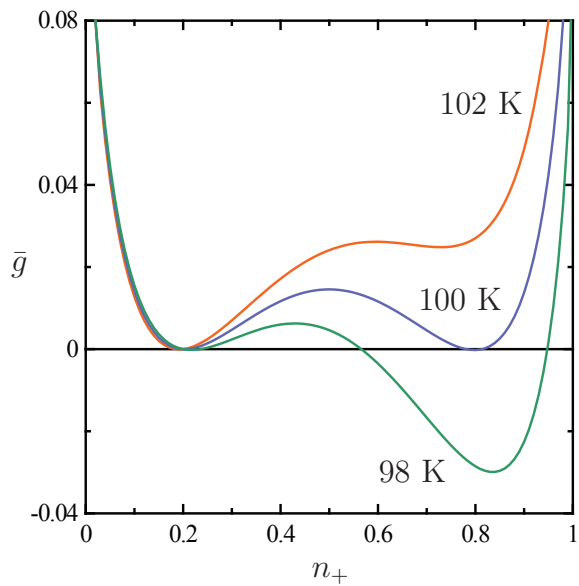}
           \caption{(Color online) Scaled Gibbs free energy per particle $\bar g\equiv G/Nk_BT$ for a system of hard spheres of diameter $\sigma=3$ \AA\ and mass $m=3\times 10^{-26}\,\,{\rm kg}$ as obtained from (1) to (5) as a function of the order parameter $n_+$ at $p=3160\,\,{\rm bar}$ and temperatures indicated explicitly. The stable phase is at 98 K the one with higher $n_+$ (solid) and at 102 K the one with lower $n_+$ (fluid). Solid-fluid coexistence occurs at 100 K at the selected $p$. This picture, characterized by a finite difference between the $n_+$ values of the coexisting phases, remains for any path crossing the coexistence curve in the $p$-$T$ plane. Note that all three curves have been shifted so as to get a common $\bar g$ value at the minimum corresponding to lower $n_+$.}
			\label{fig3}
\end{center}
\end{figure}

We finally note that the standard graphical analysis of Fig. 3 \cite{binder} indicates that the model's transition is certainly first-order without critical point. This is in contrast with an early approach to freezing with a spin-1 model displaying unseen features such as tricriticality or critical melting \cite{sivardiere} but it certainly agrees with Landau's conjectures on the nature of the transition \cite{landau}.

\emph{Simple substance.}---As a natural extension, we suppose that particles in nearest-neighbor cells experiment attractive interactions. We consider a background interaction energy $-\varepsilon_0$ for every pair of adjacent occupied cells supplemented by and extra energy $-\delta\varepsilon$ when particles have positional order. This clearly mimics a normal pair potential like the one due to Lennard-Jones and Devonshire, with positional order associated with the potential minimum. 

Equations (3) to (6) hold for this model of a ``simple substance,'' whereas (2) may be substituted by
$$
E={\scriptstyle\frac{3}{2}}n{\cal N}k_BT -6{\cal N}\varepsilon_0 n^2-6{\cal N}\delta\varepsilon n_+^2,
\eqno (9)
$$
implying that instead of (7) and (8) the following equations apply \cite{meanfield}: 
$$
\ln{n_+\over (n-n_+)\omega}=n_+[2(\lambda-{\scriptstyle\frac{\sqrt 2}{2}})\bar p+12\beta\delta\varepsilon],
\eqno (10)
$$
$$
\bar p [\lambda+(\lambda -{\scriptstyle\frac{\sqrt 2}{2}})n_+^2] =-\ln(1-n) -6\varepsilon_0 n^2-6\delta\varepsilon n_+^2.
\eqno (11)
$$ 

\begin{figure*}[!t] 
          \includegraphics[width=0.95\textwidth]{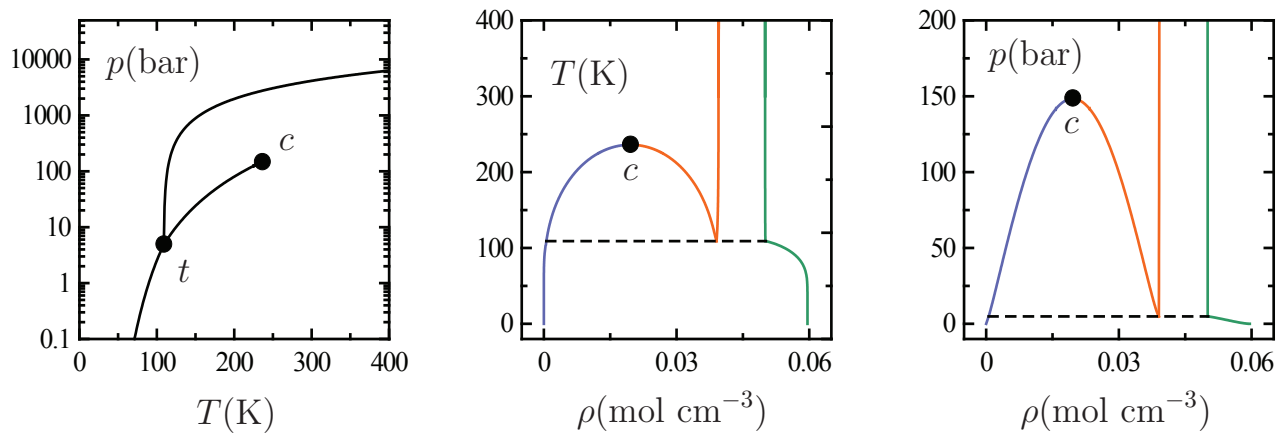}
           \caption{(Color online) Phase diagram for the simple-substance model in the $p$-$T$, $T$-$\rho$, and $p$-$\rho$ planes calculated as indicated in the text. Lines in the $p$-$T$ plane determine the conditions of two-phase coexistence bounded by the triple point $t$ and the gas-liquid critical point $c$. Lines in the $T$-$\rho$ and $p$-$\rho$ planes enclose regions of two-phase coexistence for solid (green), liquid (orange), and gas (purple), with horizontal dashed lines joining the states of three-phase coexistence associated with $t$.}
			\label{fig1}
\end{figure*}

To gauge the ability of our augmented model, we make connection with the Lennard-Jones parameters of argon, namely, $\sigma_{\rm LJ}^{\rm argon}\simeq 3.4$ \AA\ and $\varepsilon_{\rm LJ}^{\rm argon}\simeq 0.0103$ eV \cite{hansen}. Thus, with the same $\omega$ and $\lambda$ values adopted for the hard-sphere system, we set $\sigma=\sigma_{\rm LJ}^{\rm argon}$ and choose optimal $\varepsilon_0 \simeq 0.00674$ eV and $\delta\varepsilon \simeq 0.00363$ eV values fulfilling $\varepsilon_0+\delta\varepsilon = \varepsilon_{\rm LJ}$. Then, given any two among $T$, $p$, and $\rho$, (6), (10), and (11) were solved numerically for the remaining one as well as for $n$ and $n_+$. This allows every model's thermodynamic property to be obtained.

Figure 4 shows that the resulting phase diagram in the $p$-$T$, $T$-$\rho$, and $p$-$\rho$ planes meets all basic features revealed by experiment, while differences arise at a quantitative level owing to the coarse-grained nature of the model and the approximate mean-field solutions explored. Moreover, the model's gas-liquid and solid-liquid coexistence curves in the $T$-$\rho$ and $p$-$\rho$ planes reflect the symmetries of the underlying Ising model. Experimental evidence indicates that such symmetries are absent and there are indeed models describing them for the gas-liquid case (see, e.g., Ref. 10). Clearly, the task of extending our present model so as to describe asymmetry-related attributes of the solid-liquid coexistence curve (sloped diameter and low-temperature widening, mainly) is to be undertaken.

It is to be emphasized that this fairly accurate qualitative picture has been achieved by simply incorporating the main features of the attractive piece of a normal pair potential to a model accounting for the effect of primitive hard-core repulsion. This is consistent in the main with the classic work by Longuet-Higgins and Widom \cite{longuet-higginswidom} on the solid-liquid transition of simple substances, albeit their approach is different from ours inasmuch as its starting point is the corrected hard-sphere equation of state provided by molecular simulation. In any event, common to both the pioneering approach and the present one is certainly the idea that harsh repulsive forces underlie the whole behavior, with attractive interactions entering as a perturbation. This picture goes back to van der Waals original work, whose renaissance during the 1960s was spurred by the original simulations for the freezing of hard spheres \cite{alder} and stimulated the development of the Modern Liquid State Science \cite{widom,chandler}. In this connection, our present approach is reflecting that the validity of the van der Waals picture holds for the crystalline solid, for its transitions to liquid and gas, and, as a result, for the \emph{whole} phase behavior of a simple substance.

\emph{Outlook.}---It is natural to further extend the present investigation to substances beyond the van der Waals paradigm. One prominent candidate is water. Incorporation of the orientational degrees of freedom associated with hydrogen-bonding to our simple-substance model looks promising and can be done along the lines depicted previously \cite{isingparadigm}. While it is hard to imagine a sufficiently general Ising model accounting for the many existing ice phases, we do believe it possible to approach the interplay between the solid and supercooled liquid phases \cite{nilsson,lltransition,lltransition2}.

This entails metastability, which is another line of inquiry itself. The topic can be readily approached from mean-field theory \cite{binder} and of particular interest is the possible existence of a limit of supercooling suggested by the van der Waals loop in Fig. 2. A theoretical estimation of such a limit for the supercooled hard-sphere fluid phase was indeed early provided by Kirkwood \cite{kirkwood} and while this fundamental question has been analyzed from time to time for real substances \cite{brout,gallium} it is seemingly far from being settled \cite{stillinger}. We do find it worthy to revisit the problem in the near future with the aid of the simple models introduced here.

Support from the Spanish Ministry of Science and Innovation under grant no. PID2020-115722GB-C22 is greatly acknowledged.

\newpage

%
%
















\begin{thebibliography}{120}
\bibitem{comment} This is rigorously true provided $\dot v_1$ is, as we assume, vanishingly small relative to the cell volume; note also that we have used that the lattice coordination number is 12.
\bibitem{meanfield} We choose to impose $(\partial\mu/\partial n)_{T,p,n_+}=0$ and $(\partial\mu/\partial n_+)_{T,p,n}=0$, with $\mu=(E-TS+pV)/N$ obtained from the equations displayed in the text.
\bibitem{comment2} Note that cell models provide an oversimplified description of the excluded-volume problem.
\end{thebibliography}
\end{document}